\documentclass[12pt]{article}
\usepackage{a4wide,amsmath,amsthm}
\usepackage{amsmath,amsthm}

\newcommand{\url}[1]{}

\newcommand{\inverse}[1]{{\textstyle\frac{1}{#1}}}
\newcommand{\half}{\inverse{2}}

\title{{\bf \Large The general mixture-diffusion SDE and its relationship with an uncertain-volatility
option model with volatility-asset decorrelation}\thanks{Draft.
This working paper is downloadable at {\tt
http://www.damianobrigo.it}}
  }
\author{Damiano Brigo\\
Banca IMI, San Paolo IMI Group \\
Corso Matteotti 6 -- 20121 Milano, Italy \\
  Fax: +39 02 7601 9324  \\ E-mail:\ {\tt
  damiano.brigo@bancaimi.it}\\
  {\tt http://www.damianobrigo.it} \
}
\date{First Version: March 15, 2002. This Version: September 10, 2002}
\newtheorem{theorem}{Theorem}[section]
\newtheorem{proposition}[theorem]{Proposition}

\newtheorem{corollary}[theorem]{Corollary}

\newtheorem{remark}[theorem]{Remark}
\newtheorem{problem}[theorem]{Problem}

\begin{document}
\maketitle
\thispagestyle{empty}
\begin{abstract}
In the present paper, given an evolving mixture of probability
densities, we define a candidate diffusion process whose marginal
law follows the same evolution. We derive as a particular case a
stochastic differential equation (SDE) admitting a unique strong
solution and whose density evolves as a mixture of Gaussian
densities. We present an interesting result on the comparison
between the instantaneous and the terminal correlation between the
obtained process and its squared diffusion coefficient. As an
application to mathematical finance, we construct diffusion
processes whose marginal densities are mixtures of lognormal
densities. We explain how such processes can be used to model the
market smile phenomenon. We show that the lognormal mixture
dynamics is the one-dimensional diffusion version of a suitable
uncertain volatility model, and suitably reinterpret the earlier
correlation result. We explore numerically the relationship
between the future smile structures of both the diffusion and the
uncertain volatility versions.
\end{abstract}


\vspace{0.5cm}

{\bf AMS classification codes}: 60H10, 60J60, 91B28, 91B70

%
%
%
%
%
%

\tableofcontents

\newpage

\section{Introduction: SDEs and mixtures}
Let us consider the scalar stochastic differential equation (SDE)
\begin{eqnarray} \label{sde1}
   dX_t =  f_t(X_t) dt + \sigma_t(X_t) d W_t,
\end{eqnarray}
of diffusion type, with deterministic initial condition $X_0 =
x_0$, and where $\{W_t, t\ge 0\}$ is a standard Brownian motion.
We assume that
\begin{itemize}
   \item[(A1)]  The stochastic differential equation~(\ref{sde1})
      characterized by the coefficients $f$, $\sigma$,
      and by the initial condition $x_0$ admits a unique strong
      solution, whose support is assumed to be the interval $(b,\infty)$ at all
      time instants. The symbol $b$ denotes either a real number
      (typically $0$) or $-\infty$.
\end{itemize}
Under (A1), we can analyze the distribution of our SDE's solution
at all time instants. In describing the evolution of the
distribution of a diffusion process, the Fokker--Planck partial
differential equation is a fundamental tool. We assume that
\begin{itemize}
 \item[(A2)]
The unique solution $X_t$ of~(\ref{sde1}) admits a density $p_t$
that is absolutely continuous with respect to the Lebesgue measure
in $(b,\infty)$
and that satisfies the Fokker--Planck equation:
\begin{eqnarray*} 
\frac{\partial p_t}{\partial t} =  -\frac{\partial}{\partial x} (f_t p_t) + \half
   \frac{\partial^2}{\partial x^2} (a_t p_t),  \ \    a_t(\cdot) = \sigma_t^2(\cdot) \ .
\end{eqnarray*}
\end{itemize}
The other main ingredients in the present paper are mixtures of
densities. More specifically, we will consider a basic parametric
family of densities, say
\[ {\cal D} = \{ p(\cdot,\theta), \theta \in \Theta \} , \]
with $\Theta$ open in ${\bf R}^d$, being $d$ a suitable integer,
and where all densities in the family share a common support
$(b,\infty)$. We are interested in considering a particular
mixture of densities in this family. In other words, we fix a set
of non-negative weights $\lambda_1,\ldots,\lambda_m$, $\lambda \ge
0$, $\sum_i \lambda_i =1$. We take the space of all possible
mixtures of densities in ${\cal D}$ with fixed weights $\lambda$:

\[ {\cal M}({\cal D},\lambda) := \{\lambda_1 p(\cdot,\theta_1) +
\ldots +\lambda_m p(\cdot,\theta_m), \ \ \
\theta_1,\ldots,\theta_m \in \Theta\} .
\]

We will refer to this set of densities as to the $\lambda$-mixture
family for ${\cal D}$, or shortly as to the ``mixture family" when
$\lambda$ is clear from the context.
%
%
We are interested in finding an SDE whose solution $X_t$ has a
density $p_t$ that follows a prescribed evolution in a given
mixture family. More precisely, we require the curve $t \mapsto
p_t$, in the space of all densities, to match a given curve $t
\mapsto \sum_{i=1}^m \lambda_i p(\cdot,\theta_i(t))$ in a given
${\cal M}({\cal D},\lambda)$.
%
%
%
\begin{problem} \label{fin-dim:pro}
Let be given a mixture family ${\cal M}({\cal D},\lambda)$ of
densities with support $(b,\infty)$, and a drift $f_t(x)$
satisfying
\[ \mbox{(A3)} \ \ \ \lim_{y\rightarrow b+} f_t(y) p(y,\theta) = 0 \ \ \mbox{for
all} \ \ t\ge 0, \ \  \theta \in \Theta. \] Let $\Sigma(f,x_0)$
denote the set of all real-valued diffusion coefficients
$\sigma^f$ such that the related SDE~(\ref{sde1}) satisfies
assumptions (A1) and (A2), and such that
\[\mbox{(A4)}\ \ \lim_{y\rightarrow b+} \sigma^f_t(y)^2 p(y,\theta) = 0, \ \
\lim_{y\rightarrow b+} \frac{\partial}{\partial
y}\left(\sigma^f_t(y)^2 p(y,\theta)\right) = 0 \ \ \mbox{for all}
\ \ t\ge 0, \ \ \ \theta \in \Theta.
\]

Assume $\Sigma(f,x_0)$ to be non-empty. Then, given the curve $t
\mapsto \sum_{i=1}^m \lambda_i p(\cdot,\theta_i(t))$ in ${\cal
M}({\cal D},\lambda)$ (where $t \mapsto \theta_i(t)$ are
$C^1$--curves in the parameter space $\Theta$),
%
%
find a diffusion coefficient in $\Sigma(f,x_0)$ whose related SDE
has a solution with density $p_t = \sum_{i=1}^m \lambda_i
p(\cdot,\theta_i(t))$.
\end{problem}

This problem is the analogous of the drift search problem
described and solved in Brigo and Mercurio (1998) and Brigo
(2000). The solution of this problem is given by the following.
\begin{proposition} \label{sol-prob1}
{\bf (Solution of Problem~\ref{fin-dim:pro})}
Assumptions and notation of Problem~\ref{fin-dim:pro} in force.
Consider the stochastic differential equation
\begin{eqnarray} \label{sol:prob1}
d Y_t &=& f_t(Y_t) dt + \sigma^f_t(Y_t) dW_t, \ \ Y_0 = x_0,
\\ \nonumber
(\sigma^f_t(y))^2 &=& \frac{2}{\sum_{i=1}^m \lambda_i
p(y,\theta_i(t))} \left[ \int_{b}^y \left(\int_{b}^x \sum_{i=1}^m
\lambda_i \frac{\partial p(u,\theta_i(t))}{\partial t} du \right)
dx \right. \\ \nonumber  & & \ \ \ \ \left. + \int_{b}^y f_t(x)
\sum_{i=1}^m \lambda_i p(x,\theta_i(t)) dx \right].
\end{eqnarray}
%
If  $\sigma^f \in \Sigma(f,x_0)$, then the SDE (\ref{sol:prob1})
solves Problem~\ref{fin-dim:pro}, in that
\begin{displaymath}
p_{Y_t}(y) = \sum_{i=1}^m \lambda_i\ p(y,\theta_i(t)), \ \ t \ge
0.
\end{displaymath}
\end{proposition}
\begin{proof}
Write the Fokker--Planck equation for the candidate diffusion $Y$
with the candidate solution $\sum_{i=1}^m \lambda_i
p(\cdot,\theta_i(t))$ already inside, and then back-out the
diffusion coefficient via two subsequent integrations starting
from the lower point $b$.
\end{proof}
We will use a particular case of this proposition.
\begin{corollary} Assumptions as in the above proposition. If the
basic densities $p(\cdot,\theta_i(t))$  evolving in ${\cal
M}({\cal D},\lambda)$ are the marginal densities of a family of
(instrumental) SDEs
\[ d X^i_t = f^i_t(X^i_t) dt + \sigma^i_t(X^i_t) dW_t, \ \ x_0,  \ \ \
p(x,\theta_i(t)) := p_{X^i_t}(x) , \] all satisfying assumptions
(A1), (A2), (A3) (with $f^i$'s replacing $f$) and such that
$\sigma^i \in \Sigma(f^i,x_0)$,  then the solution of
Problem~(\ref{fin-dim:pro}) takes the form
\begin{eqnarray}\label{sigmamixture} (\sigma^f_t(y))^2 = \sum_{i=1}^m
\Lambda^i_t(y) (\sigma^i_t(y))^2 +  \frac{2\sum_{i=1}^m \lambda_i
\int_{b}^y (f_t(x)-f^i_t(x)) p(x,\theta_i(t)) dx }{\sum_{j=1}^m
\lambda_j p(y,\theta_j(t))} , \\ \nonumber \Lambda^i_t(y) :=
\frac{\lambda_i p(y,\theta_i(t))}{\sum_{j=1}^m \lambda_j
p(y,\theta_j(t))} \ \ \  (\mbox{with} \ \  \sum_{j=1}^m
\Lambda^i_t(y) = 1).
\end{eqnarray} An interesting particular case occurs when $f$
satisfies
\begin{equation}\label{driftmixture} f_t(y) = \sum_{i=1}^m \Lambda^i_t(y)
f^i_t(y).
\end{equation}
In this case the second term in the right hand side
of~(\ref{sigmamixture}) vanishes, and we have that requiring the
marginal of our final SDE to be a given $\lambda$-mixture of the
marginals of the instrumental processes results in a drift and
squared diffusion coefficients that are (state-dependent)
$\Lambda$-``mixtures" of the drifts and squared diffusion
coefficients of the instrumental processes. We move from the
combinators $\lambda$ (density) to the combinators $\Lambda$
(coefficients).
\end{corollary}

\begin{proof} It suffices to use the previous result by noticing that the
Fokker--Planck equation for the instrumental processes reads:
\[ \frac{\partial p(x,\theta_i(t))}{\partial t} = -\frac{\partial[f^i_t(x)p(x,\theta_i(t))]}{\partial x}
+ \half \frac{\partial^2 [ \sigma^i_t(x)^2
p(x,\theta_i(t))]}{\partial x^2} \] and substituting
in~(\ref{sol:prob1}) the right hand sides of such Fokker--Planck
equations.
\end{proof}

The problem solved in the above proposition could have been
formulated to track a generic density-evolution $t \mapsto q_t$
which does not necessarily occur in a mixture family. In such a
case, given assumptions analogous to (A3) and (A4), the
appropriate diffusion coefficient would be
\begin{eqnarray*}
(\sigma^f_t(y))^2 = \frac{2}{q_t(y)} \left[ \int_{b}^y
\left(\int_{b}^x \ \frac{\partial q_t(u)}{\partial t} du \right)
dx + \int_{b}^y f_t(x) q_t(x) dx \right]
\end{eqnarray*}
which generalizes the solution in the above proposition. We
preferred to present directly the parametric case.
%
%
In the next sections, we shall consider first an interesting
application of the above corollary to the fundamental cases of
mixtures of normal and lognormal families, and second an
application of the latter to the option pricing problem in
mathematical finance, in particular as a possible means to model
the so called ``smile" phenomenon.
%
%
%
%
\section{Diffusions whose densities follow mixtures of normal
distributions}\label{normalmixturesection} Mixtures of normals are
ubiquitous in statistics, representing the standard in many
applications. Also, mixture of normals are often used in
econometrics to model time series that have tails fatter than the
Gaussian. See for example Alexander~(2001) for a discussion on
normal mixtures applied to financial data, also in comparison with
different distributions. In general, normal mixtures represent in
a sense the least departure from the Gaussian family allowing for
skewness and kurtosis different from the Gaussian ones. However,
to the best of our knowledge, no explicit attempt has been made so
far to design continuous time diffusion models displaying this
kind of marginal distributions. The subject can be relevant at
least in mathematical finance, as we will show when addressing the
smile problem, but bears also an interest of its own in the study
of the interaction between a diffusion process dynamics and
particular families of distributions.

Let us then start from the normal family, that we parameterize via
its first two moments. In this case
\[ {\cal D} = \{ p_{{\cal N}(m,v^2)}, \ \ m,v \in {\bf R}\}, \ \ \ b = -\infty. \]
We are given a curve in the $\lambda$-mixtures of normals, $t
\mapsto \sum_{i=1}^m \lambda_i p_{{\cal N}(m_i(t),v_i(t)^2)}$, and
we wish to find a diffusion process compatible with such a law. It
is easy to see that if we consider the instrumental processes
\[ d X^i_t = \mu_i(t) \ dt + \sigma^i(t) d W_t, \int_0^t \mu_i(s) ds = m_i(t), \int_0^t (\sigma^i(s))^2 ds = v_i(t)^2  \]
with deterministic $\mu$ and $\sigma$'s and null initial
condition, by applying the above corollary  in the particular
case~(\ref{driftmixture}) we end up with the diffusion process
whose drift and squared diffusion coefficient are given
respectively by

\begin{eqnarray}\label{mixtureofnormals}
f_t(y) = \sum_{i=1}^m \Lambda^i_t(y) \mu^i(t), \ \ \ \
\sigma^f_t(y)^2 = \sum_{i=1}^m \Lambda^i_t(y) \sigma^i(t)^2 , \ \
\\ \nonumber \ \Lambda^i_t(y) = \frac{\lambda_i p_{{\cal
N}(m_i(t),v_i(t)^2)}(y)}{\sum_{j=1}^m \lambda_j p_{{\cal
N}(m_j(t),v_j(t)^2)}(y)} .
\end{eqnarray}

Now assume the coefficients $t \mapsto \mu(t)$'s and $t \mapsto
\sigma(t)$'s to be at least $C^1$ (and hence bounded on all finite
time intervals), with the $\sigma$'s bounded away from zero,
$\sigma_i(t) \ge L > 0$ for all $i$ and $t$. There are possible
problems for a regular behaviour of the above $f$ and $\sigma^f$
when  $(t,y) \rightarrow (0,0)$. In order to avoid this, we may
decide to modify our coefficients by imposing
$\mu_i(t)=\bar{\mu}$, $\sigma_i(t) = \bar{\sigma}$ for all $i$ and
$t\in [0,\epsilon)$, with $\epsilon$ a given positive real number,
typically small. We can then assume suitable transitory
trajectories for the $\mu$'s and the $\sigma$'s in say $[\epsilon,
2\epsilon)$ that recover the correct integrals $m$ and $v^2$ for
times larger than $2\epsilon$. The desired mixture is thus
unmatched only in the (typically negligible) time interval $[0, 2
\epsilon]$. Now $\Lambda^i_t(y)=\lambda_i$ for all $t < \epsilon$
and all $y$.

From the above assumptions, since by definition $0\le
\Lambda^i_t(y) \le 1$ for all $t$ and $y$, we have also that $f$
and $\sigma^f$ are bounded in all regions with bounded time,
implying non-explosion for the related SDE in all intervals of the
kind $t \le T$ for some $T>0$. Furthermore, $f$ and $(\sigma^f)^2$
are also seen to be $C^1$ in both $t$ and $y$. This in turn
implies that $f$ is locally Lipschitz, whereas the fact that the
$\sigma(t)$'s are bounded away from zero implies that so is
$\sigma^f$ in both $t$ and $y$, thus ensuring that $(\sigma^f)^2
\in C^1$ implies $\sigma^f \in C^1$. Therefore $\sigma^f$ is
locally Lipschitz, and now we have all the elements to apply
Theorem 12.1 in Section V.12 of Rogers and Williams (1996) to
conclude that our SDE admits a unique strong solution. We have
thus produced a diffusion process compatible with a given mixture
of normal densities, and can state the following
\begin{theorem} {\bf (SDEs whose marginal law follows a given normal mixture)}
Consider an SDE
\[ d Y_t = f_t(Y_t) dt +  \sigma^f_t(Y_t) d W_t , \ \ Y_0 = 0 ,  \]
with drift and diffusion coefficient $f$ and $\sigma^f$ as
in~(\ref{mixtureofnormals}), where the $m$'s and $v$'s are $C^2$
time functions. Assume moreover that in an initial time interval
$t\in [0,\epsilon)$ we have $m_i(t) = \bar{\mu} t$ and
$v_i(t)^2=\bar{\sigma}^2 t$. Then the considered SDE admits a
unique strong solution and its solution has as marginal density
the $\lambda$-normal mixture
\[ p_{Y_t} =  \sum_{i=1}^m \lambda_i p_{{\cal N}(m_i(t),v_i(t)^2)} \]
at time $t$.
\end{theorem}

An interesting feature of the obtained process concerns the
covariance between the process itself and the diffusion
coefficient in its dynamics. This quantity is of interest, for
example, in mathematical finance. Let us denote by
``$\mbox{corr}_t$" the correlation between two random variables,
and by ``$\mbox{cov}_t$" the covariance, both conditional on the
information available at time $t$, the time being omitted if
$t=0$. We have, for the above SDE, the following {\em
``instantaneous"} correlation between the {\em instantaneous
change} in the process and the {\em instantaneous change} in the
related diffusion coefficient at a given instant:

\[ \mbox{corr}_t(dY_t,d\  \sigma^f_t(Y_t)) =
\frac{d \langle Y, \sigma^f_\cdot(Y_\cdot )\rangle_t }{\sqrt{d
\langle Y \rangle_t\ \ d \langle \sigma^f_\cdot(Y_\cdot)
\rangle_t}} = \frac{d Y_t \ d\ \sigma^f_t(Y_t)}{\sqrt{d Y_t \ d
Y_t} \ \sqrt{ d\  \sigma^f_t(Y_t) \ d \ \sigma^f_t(Y_t) }} = 1 ,
\] as is obvious from the fact that the diffusion coefficient is a
deterministic function of the current value of $Y$.

However, things are rather different for the {\em terminal}
correlation. A straightforward if lengthy computation is needed to
show that

\[ \mbox{cov} (Y_t,\sigma^f_t(Y_t)^2)= E_0( Y_t \  \sigma^f_t(Y_t)^2)
- E_0( Y_t) \  E_0(\sigma^f_t(Y_t)^2) \]\[= \sum_{i=1}^m \lambda_i
m_i(t) \sigma_i^2(t) - \left(\sum_{i=1}^m \lambda_i
m_i(t)\right)\left( \sum_{i=1}^m \lambda_i \sigma_i^2(t)\right) \
.
\]

Consider now the case where all the means in the normal mixture
densities are equal: $\mu_i(\cdot) = \mu(\cdot)$ for all $i$, and
correspondingly $m_i(\cdot) = m(\cdot)$. In this case $f_t(y) =
\mu(t)$ for all $y$ and, perhaps surprisingly, especially if
compared to the perfect instantaneous correlation, the above
formula gives

\[ \mbox{corr}(Y_t,\sigma^\mu_t(Y_t)^2) = \mbox{cov}(Y_t,\sigma^\mu_t(Y_t)^2) = 0  . \]

This is a case where the {\em instantaneous} correlation is $1$,
whereas the terminal correlation after a time $t$, no matter how
small, is $0$. Thus we have a stochastic process whose
instantaneous {\em changes} are perfectly correlated with the
instantaneous {\em changes} of its squared diffusion coefficient,
whereas at any time its value has 0 correlation with the squared
diffusion coefficient value. It seems then that the squared
diffusion coefficient has a special shape that immediately
``decorrelates" itself from the process even after an arbitrarily
small time. It is not difficult to prove an analogous statement
for the {\em average} squared diffusion coefficient:

\[ \mbox{corr}\left(Y_T,\int_0^T \sigma^\mu_t(Y_t)^2 \ dt \right) = 0  . \]

Such interesting results on terminal correlation versus
instantaneous correlation will be further discussed in the
financial applications.

Finally, going back to the above theorem, we notice that by
reasoning along the same lines, diffusions displaying mixtures of
lognormals as marginal densities can be easily obtained. Indeed,
if $Y$ is the diffusion process from the above theorem, we can set
$S_t := \exp(Y_t)$ and derive easily the SDE for $S$ via Ito's
formula:
\[ d S_t = S_t f_t(\ln(S_t)) dt + \half S_t  \sigma^f_t(\ln(S_t))^2 dt + S_t \sigma^f_t(\ln(S_t)) d
W_t , \ \ S_0 = 1. \]

 It is a straightforward exercise to verify that the
equality
\[ p_{S_t}(y) = p_{Y_t}(\ln y)/y \]
implies that $S$ is distributed as a mixture of lognormals if $Y$
is distributed as a mixture of normals.

\section{The smile phenomenon in option pricing}

Now we briefly review a stylized version of the smile problem in
financial modeling and explain the possible use of diffusions
whose marginals follow $\lambda$-mixtures of lognormals.

\subsection{The smile problem and market implied distributions}
Let us consider a financial market with a ``money market account"
process $B_t$, with positive deterministic instantaneous interest
rate $r(t)>0$, so that $dB_t=r(t) B_t dt$. Let us also consider a
process $S_t$ modeling the evolution of some traded financial
(risky) asset in our market, typically a stock.

The resulting financial market might admit arbitrage
opportunities. A sufficient condition which ensures arbitrage-free
dynamics is the existence of an equivalent martingale measure~$Q$,
sometimes termed risk-neutral measure. An equivalent martingale
measure is a probability measure that is equivalent to the initial
one and under which the process $\{S_t/B_t:t\ge 0\}$ is a
martingale. Let us assume, in line with the basic Black and
Scholes (1973) setup, that the risk-neutral dynamics of $S$ is
modeled by
\begin{eqnarray} \label{BeS}
 d S_t = r(t) S_t dt + \nu(t)  S_t \ dW_t, \ \  S_0 = s_0 , \ \
t \in [0,T],
\end{eqnarray}
where $s_0$ is a positive deterministic initial condition, and
$\nu$ is a well-behaving deterministic function of time
(instantaneous volatility). The above process is a geometric
Brownian motion and the probability density $p_{S_t}$ of $S_t$, at
any time $t$, is lognormal. Indeed,
\begin{eqnarray} \label{BeSret}
 \ln \frac{S_t}{S_0}
\sim {\cal N}\left( R(0,t) - \half V(t)^2 ,  \ V(t)^2  \right), \
R(a,t) := \int_a^t  r(s)  ds , \ V(t)^2 := \int_0^t \nu(s)^2 ds .
\end{eqnarray}
When $a=0$, we write shortly $R(t)$ for $R(0,t)$. In the above
equation~(\ref{BeS}) we modeled directly the risky asset dynamics
under the unique equivalent martingale measure, so that $W$ is
assumed to be a Brownian motion under that measure, and it is
immediate to check that $S_t/B_t$ is indeed a martingale. It is
this dynamics that matters when pricing options, as opposed to the
real world one, which is related instead to historical estimation,
statistical analysis and similar matters. Indeed, by applying the
results by Harrison and Pliska (1981), the unique no-arbitrage
price for a given ${\cal F}_T$-measurable contingent claim $H_T\in
L^2(Q)$ is $V_t=B_tE^{Q}\left\{\left.H_T/B_T\right| {\cal
F}_t\right\}= : B_tE_t^{Q}\{H_T/B_T\} $ where $\{{\cal F}_t:t\ge
0\}$ denotes the filtration associated to the process $S$. The
contingent claim is said to be a simple one when it is of the form
$H_T = h(S_T)$ for a suitable function $h$.

One of the most common simple claims is a European call option
written on the stock, with maturity $T$ and strike $K$, which pays
$H=(S_T-K)^+$ at time $T$. Its price is obtained by computing the
expectation of the discounted payoff according to the lognormal
distribution implied by~(\ref{BeSret}), leading to the celebrated
Black and Scholes (1973) call option formula, which we denote by
``BSCall" and whose explicit expression we omit for brevity:
\[ E^Q_0[(S_T-K)^+/B(T)] = \mbox{BSCall}(S_0,K,T,R(T),V(T)) . \]
The quantity $V(T)/\sqrt{T}$ is the (average) volatility of the
option, and according to this formulation, does not depend on the
strike $K$ of the option. Indeed, in this formulation, volatility
is a characteristic of the stock $S$ underlying the contract, and
has nothing to do with the nature of the contract itself. In
particular, it has nothing to do with the strike $K$ of the
option.


Now take two different strikes $K_1$ and $K_2$. Suppose that the
market provides us with the prices of two related options on our
stock with the same maturity $T$: $\mbox{MKTCall}(S_0,K_1,T)$ and
$\mbox{MKTCall}(S_0,K_2,T)$.

Life would be simple if the market followed Black and Scholes'
formula in a consistent way. But is this the case? Does there
exist a {\em single} volatility parameter $V(T)$ such that both
the following equations hold?
\[ \mbox{MKTCall}(S_0,K_1,T) =
\mbox{BSCall}(S_0,K_1,T,R(T),V(T)),\]\[ \mbox{MKTCall}(S_0,K_2,T)
= \mbox{BSCall}(S_0,K_2,T,R(T),V(T)).
\]


 The answer is a resounding ``no''. In general, market option
prices do not behave like this. What one sees when looking at the
market is that two {\em different} volatilities $V(T,K_1)$ and
$V(T,K_2)$ are required to match the observed market prices if one
is to use Black and Scholes' formula:
\[ \mbox{MKTCall}(S_0,K_1,T) =
\mbox{BSCall}(S_0,K_1,T,R(T),V(T,K_1)),\]\[
\mbox{MKTCall}(S_0,K_2,T) =
\mbox{BSCall}(S_0,K_2,T,R(T),V(T,K_2)).
\]

In other terms, each market option price requires its own Black
and Scholes (implied) volatility $V^{\mbox{\tiny
MKT}}(T,K)/\sqrt{T}$ depending on the option strike $K$.

The market therefore uses Black and Scholes' formula simply as a
metric to express option prices as volatilities. The curve $K
\mapsto V^{\mbox{\tiny MKT}}(T,K)/\sqrt{T}$ is the so called
volatility smile of the $T$-maturity option. If Black and Scholes'
model were consistent along different strikes, this curve would be
flat, since volatility should not depend on the strike $K$.
Instead, this curve is commonly seen to exhibit ``smiley" or
``skewed'' shapes.

Clearly, only some strikes $K=K_i$ and maturities $T=T_j$ are
quoted by the market, so that usually the remaining points have to
be determined through interpolation or through an alternative
model. Interpolation in $K$, for a fixed maturity $T$, can be easy
but it does not give any insight as to the underlying stock
dynamics compatible with such prices.

Indeed, suppose that we have a few market option prices for
expiries $T=T_j$ and for a set of strikes $K=K_i$.

For each fixed $T=T_j$, by smooth interpolation we can obtain the
price for every other possible $K$, i.e. we can build a function
$K \mapsto \mbox{MKTCall}(S_0,K,T)$. Now, if this strike-$K$ price
corresponds really to an expectation, we have
\begin{eqnarray}
\mbox{MKTCall}(S_0,K,T) =
 e^{-\int_0^T r(s) ds}  E^Q_0 (S_T - K)^+ = e^{-\int_0^T r(s) ds}  \int_K^\infty (x-K) p_{T}(x)\, dx
 ,
\end{eqnarray}
where $p_T$ is the true risk-neutral density of the underlying
stock at time $T$. If Black and Scholes' formula were consistent,
this density would be the lognormal density, coming for example
from a dynamics such as~(\ref{BeS}), i.e. $p_{T}=p_{S_T}$. We have
seen that this is not the case in the market. However, by
differentiating the above integral twice with respect to $K$ we
see that, see also Breeden and Litzenberger (1978),
\[ \frac{\partial^2 \mbox{MKTCall}(S_0,K,T)}{\partial
K^2}=  e^{-\int_0^T r(s) ds} p_{T}(K) , \]
so that by differentiating the interpolated-prices curve we can
find the density $p_T$ of the underlying stock at time $T$ that is
compatible with the given interpolated prices. Nevertheless, the
method of interpolation may interfere with the recovery of the
density, since a second derivative of the interpolated curve is
involved. Moreover, what kind of dynamics, alternative
to~(\ref{BeS}), do the densities
$p_{T_1},p_{T_2},\ldots,p_{T_j},\ldots$ come from?

\subsection{Local and stochastic volatility models}

A partial answer to these issues can be given the other way
around, by starting from an alternative dynamics. Indeed, assume
that
\begin{eqnarray} \label{BeSaltern}
 d S_t = r(t) S_t dt +   \sigma(t,S_t)\ S_t \ dW_t, \ \  S_0 = s_0 , \ \
\end{eqnarray}
where $\sigma$ can be either a deterministic or a stochastic
function of $S_t$. In the latter case we would be using a so
called ``stochastic-volatility model", where for example
$\sigma(t,S) = \xi(t)$, with $\xi$ following a second stochastic
differential equation, such as:
\[ d (\xi(t)^2) = b(t,\xi(t)^2) dt + \chi(t,\xi(t)^2) d Z_t, \ \
\]
with the important specification
\[ d Z_t d W_t = \rho\ dt  . \]
Instead, in the so-called ``local volatility models" the diffusion
coefficient $\sigma(t,S_t)$ is a deterministic function of $S_t$.

One feature of stochastic volatility models that is usually deemed
to render them superior with respect to local volatility models is
``instantaneous decorrelation". Indeed, for stochastic volatility
models we can have
\[\mbox{Corr}(d S_t, \ d \sigma^2(t,S_t)) = \rho < 1 \]
whereas
\[\mbox{Corr}(d S_t, \ d \sigma^2(t,S_t)) = 1 \]
for local volatility models, including our mixture dynamics
models. This superiority no longer holds when considering terminal
correlations, as we will remark later on. For the time being we
concentrate on deterministic $\sigma(t,\cdot)$'s, leading to
local-volatility models, such as for example $\sigma(t,S) = \eta \
S^\gamma$ (CEV model, see Cox~(1975)), where $\gamma$ ranges in a
suitable interval and where $\eta$ is a positive deterministic
constant. Below we will propose a new $\sigma(t,\cdot)$ of our
own, flexible enough for practical purposes.

We have seen above how the ``true" risk-neutral densities
$p_{T_1},p_{T_2},\ldots,p_{T_j},\ldots$ of the underlying asset
are linked to market option prices through second-order
differentiation. The problem we will face is finding a dynamics
alternative to~(\ref{BeS}) and as compatible as possible with the
densities $p_{T_1},p_{T_2},\ldots,p_{T_j},\ldots$ ideally
associated with market prices. This will be done by fitting
directly the prices implied by our alternative model to the market
prices $\mbox{MKTCall}(S_0,K,T)$ for the considered set of strikes
$K_i$ and maturities $T_j$. To further clarify this point, it may
be helpful to explain explicitly how an alternative dynamics such
as~(\ref{BeSaltern}) leads to a volatility smile to be fitted to
the market smile. The way in which an alternative local-volatility
model dynamics generates a smile is clarified by the following
stylized operational scheme:
\begin{enumerate}
\item Set the pair $(T,K)$ to a starting value;
\item Compute the model option price
\[ \Pi(T,K) = e^{-\int_0^T r(s) ds}  E^Q_0 (S_T - K)^+ \]
with $S$ obtained through the no-arbitrage alternative
dynamics~(\ref{BeSaltern}).
\item Invert Black and Scholes' formula  for this strike and maturity, i.e. solve
\[ \Pi(T,K) = \mbox{BSCall}(S_0,K,T,R(T),V(T,K)) \]
in $V(T,K)$, thus obtaining the model implied volatility $V(T,K)$.
\item Change $(T,K)$ and restart from point~2 until the last
maturity/strike pair $(T,K)$ is reached.
\end{enumerate}
The fact that the alternative dynamics is not lognormal implies
that we obtain curves in the strike $K \mapsto V(T,K)$ that are
not flat. Clearly, one needs to choose $\sigma(t,\cdot)$ flexible
enough for the surface $(T,K) \mapsto V(T,K)/\sqrt{T}$  to be able
to resemble or even match the corresponding volatility surfaces
coming from the market. Indeed, the model implied volatilities
$V(T_j,K_i)/\sqrt{T_j}$ corresponding to the observed strikes and
maturities have to be made as close as possible to the
corresponding market implied volatilities $V^{\mbox{\tiny
MKT}}(T_j,K_i)/\sqrt{T_j}$, by acting on the coefficient
$\sigma(\cdot,S)$ in the alternative dynamics.

At this point it should be clear why a $\lambda$-mixture dynamics
can be of help. Existing local volatility models have either too
little flexibility to calibrate a large number of points in a
volatility surface, or are specified in a too general way
requiring interpolation and other possibly dangerous artifices in
order to be implemented. The CEV model for example has only one
more parameter $\gamma$ with respect to the basic Black and
Scholes ``flat" model in the time-homogeneous case, so that its
fitting capabilities are rather poor. Dupire's~(1997) approach is
quite general, but if applied straightforwardly in its most
general form, it requires a continuum of traded strikes and
maturities, with the possible interpolation problems observed
above, not to mention the lack of guarantees on existence of
solutions for the resulting SDE.

\subsection{Local volatility lognormal mixture diffusion dynamics}
Consider instead our approach and write the $\lambda$-lognormal
mixture diffusion. Set the instrumental processes to Black and
Scholes processes,
\[ d X^i_t = r(t) X^i_t dt + \nu_i(t) X^i_t dW_t,  \ \ s_0
\]
and derive the diffusion coefficient corresponding to
$f(t,y)=r(t)y$. Notice that here all instrumental processes have
the same drift $f_i = f$ as the final process. In this particular
case, taking into account the lognormal marginal distributions of
the instrumental processes, by applying~(\ref{sigmamixture})
under~(\ref{driftmixture}) we have
\[ \sigma_{\mbox{\tiny mix}}^2(t,y) y^2 :=
\sigma^f_t(y)^2 = y^2 \sum_{i=1}^m \Lambda_i(t,y) \nu_i(t)^2, \]
\[ \Lambda_i(t,y)= \frac{\lambda_i\  p_{{\cal N}(\ln s_0 + R(t)-
V_i(t)^2/2,\ V_i(t)^2 )}(\ln y)} {\sum_{j=1}^m \lambda_j\ p_{{\cal
N}(\ln s_0 + R(t)- V_j(t)^2/2,\  V_j(t)^2 )}(\ln y)}, \] where $R$
and $V$'s are defined as in~(\ref{BeSret}). In Brigo and Mercurio
(2001b) we show that the SDE resulting from such coefficients,
i.e.
\begin{equation}\label{Ssdemix} d S_t = r(t) S_t\ dt + \sigma_{\mbox{\tiny mix}}(t,S_t) \ S_t \ dW_t , \ \ s_0
\end{equation}
admits a unique strong solution, provided one takes suitable
regularity conditions on the time functions $t \mapsto \nu_i(t)$'s
analogous to those on the $t \mapsto \sigma^i(t)$'s illustrated in
Section~\ref{normalmixturesection}. Indeed, it is straightforward
to prove such existence and uniqueness result starting from the
proof given in Section~\ref{normalmixturesection} for the
normal-mixture case.

For our process $S$ in~(\ref{Ssdemix}), we confirm the curious
result obtained in Section~\ref{normalmixturesection} from the
comparison between instantaneous and terminal correlations. Notice
that also in this case we have
\[\mbox{corr}_t(d S_t, \ d \sigma^2_{\mbox{\tiny mix}}(t,S_t)) = 1 ,\]
considered to be a drawback of local volatility models. Yet, when
considering terminal correlations, things change considerably.
\begin{theorem}\label{thterminalcorr} {\bf (Terminal correlation between underlying
asset and average percentage variance in the lognormal mixture
dynamics model for the smile)}

Consider the random variable
\[ v(T) := \int_0^T \sigma_{\mbox{\tiny mix}}^2(t,S_t) dt \ , \]
$v(T)/T$ being the ``average percentage variance" of the process
$S$. Then
\begin{equation}
\boxed{\mbox{corr}(\sigma^2_{\mbox{\tiny mix}}(T,S_T), S_T) = 0, \
\ \mbox{and} \ \ \mbox{corr}(v(T), S_T) = 0 \ \ \mbox{for all} \ \
T }\ .
\end{equation}
\end{theorem}
\begin{proof}
First we show that
\[ \mbox{corr}(\sigma^2_{\mbox{\tiny mix}}(T,S_T), S_T) = 0 , \]
i.e. that
\begin{equation}\label{proof1} E\{ \sigma^2_{\mbox{\tiny mix}}(T,S_T) \ S_T\} - E\{
\sigma^2_{\mbox{\tiny mix}}(T,S_T)\}E\{S_T\}= 0.
\end{equation}
This is immediate by direct calculation:
\begin{eqnarray*} E\{ \sigma^2_{\mbox{\tiny mix}}(T,S_T) \ S_T\} = \int \sum_{i=1}^m
\Lambda_i(T,y) \nu_i(T)^2  y \ p_{S_T}(y) dy = \\
 \int \sum_{i=1}^m
\Lambda_i(T,y) \nu_i(T)^2  y \ \sum_{j=1}^m \lambda_j p_{X^j_T}(y)
dy =
 \int \sum_{i=1}^m \lambda_i p_{X^j_T}(y) \nu_i(T)^2  y \ dy \\
 = s_0 e^{\int_0^T r(s)ds} \sum_{i=1}^m \lambda_i \nu_i(T)^2
\end{eqnarray*}
given the definition of the $\Lambda$'s.
Similarly, one computes
\[  E\{ \sigma^2_{\mbox{\tiny mix}}(T,S_T)\} = \sum_{i=1}^m \lambda_i
\nu_i(T)^2, \]
from which~(\ref{proof1}) follows.

To show the other equality, notice that
\[ d v(t) = \sigma^2_{\mbox{\tiny mix}}(t,S_t) dt , \]
and compute
\[ d (v(t) S_t) = S_t \sigma^2_{\mbox{\tiny mix}}(t,S_t) dt + r(t) v(t) S_t
dt  + (\ldots) d W_t . \]
Taking expectations and Fubini's theorem
\[ d E(v(t) S_t) = E(S_t \sigma^2_{\mbox{\tiny mix}}(t,S_t)) dt + r E(v(t)
S_t) dt   . \]
Set $A_t = E(S_t \sigma^2_{\mbox{\tiny mix}}(t,S_t))$, which we
computed above, and $C_t =  E(v(t) S_t)$.
The above equation reads
\[ \dot{C}_t = r(t) C_t + A_t, \]
whose solution is
\[ C_t = e^{\int_0^t r(s)ds} \int_0^t e^{-\int_0^u r(s)ds} A_u du . \]
By carrying out the computations one obtains $E(v(t) S_t)$. At
this point it is easy to prove that
\[ E(v(t) S_t) - E(v(t)) E( S_t) = 0 \]
by computing  $E(v(t))$ trough Fubini's theorem.
\end{proof}

The above result is partly weakened by the fact that correlation
is not a satisfactory measure of dependence outside the Gaussian
world. However, a striking feature remains of two processes that
are instantaneously perfectly correlated but such that for any
infinitesimal time $T = \epsilon$ have zero terminal correlation.

Let us now set apart this correlation result and go back to
understanding the reason why a $\lambda$-mixture dynamics is
particularly appealing when pricing options. One of the main
reasons lies in the price becoming a linear combination of prices
with respect to underlying assets modeled according to the
instrumental processes. Indeed, one has immediately, for a call
option,
\begin{eqnarray*}
\Pi(T,K) =  e^{-\int_0^T r(s) ds} E^Q\left\{(S_T-K)^+ \right\} =
e^{-\int_0^T r(s) ds} \int_0^{+\infty}(y-K)^+\sum_{i=1}^m\lambda_i
p_{X^i_T}(y) dy \\ =\sum_{i=1}^m \lambda_i e^{-\int_0^T r(s) ds}
\int_0^\infty (y-K)^+ p_{X^i_T}(y)dy=\sum_{i=1}^m \lambda_i \
\mbox{BSCall}(S_0,K,T,R(T),V_i(T)) \ ,
\end{eqnarray*}
the last equality following from the geometric Brownian motion
structure of the underlying instrumental processes. This procedure
is very general, and the price of a European-style simple claim is
always the linear combination of the corresponding prices for the
instrumental processes. In the lognormal-mixture case, when
pricing a call option we obtain a linear combination of Black and
Scholes prices. This is very appreciated by traders, who usually
prefer to contain departures from the lognormal distribution and
the corresponding Black and Scholes formula. In a sense, when in
need of generalizing a lognormal distribution, a mixture of
lognormals is the least departure from the original lognormal
paradigm.

Important calibration benefits that should not go unnoticed are a
consequence of the fact that a $\lambda$ mixture-of-lognormals
dynamics can price call options analytically. This is very helpful
for calibrating the model to the market. In such a case one runs
an optimization to find the values of the parameters $V_i$ and
$\lambda_i$ that best reproduce a given set of market prices, and
the target function of this optimization can be computed in closed
form without resorting to numerical methods such as Monte Carlo
simulation, trees, and finite difference schemes. Notice also that
since we are free to select an arbitrary number $m$ of
instrumental processes, in principle our diffusion model features
a limitless number of calibrating parameters. Once the model has
been calibrated, one can use it to price more complicated (for
example early exercise or path dependent) claims that have no
quoted price in the market. Monte Carlo simulation through the
Euler or Milstein discretization schemes (see for example Kl\"oden
and Platen~(1995)) applied to our dynamics or recombining trees in
the spirit of Nelson and Ramaswamy (1990) can be attempted, thanks
to the explicit diffusion dynamics we provided.

In the present paper we have presented the mixture diffusion model
in its most mathematical aspects. Other advantages and
characteristics of these models and of their variants, based on
shifted dynamics, and numerical investigation and calibrations to
market data have been illustrated in Brigo and
Mercurio~(2000a, 2000b, 2001a, 2001b).

The introduction of  general drift rates $\mu_i(t)$ not
necessarily all equal to $r$ in the instrumental processes and
possible mixtures of densities coming from hyperbolic-sine
processes are considered in Brigo, Mercurio and Sartorelli (2003).

A study of particular forms of time dependence of the $\nu$'s
leading to desirable properties of the lognormal mixture dynamics
and to a simple specification of the parameters in the model is
carried out in Alexander and Brintalos~(2003).

A generalization of the mixture dynamics apparatus to multivariate
underlying assets and possible applications to the pricing of
basket options in presence of volatility smile are considered in
Rapisarda (2002) and Brigo, Mercurio and Rapisarda (2002).

\subsection{Uncertain volatility geometric Brownian motion}

We conclude the paper by pointing out an important relationship
between the lognormal-mixture diffusion dynamics and an analogous
uncertain volatility model given by a geometric Brownian motion
with uncertain volatility.

In general it is known (Derman, Kani and Kamal~(1997),
Britten-Jones and Neuberger~(2000), Gatheral~(2001))  that every
stochastic volatility model has a local volatility (i.e.
scalar-diffusion) version that features the same marginal
distributions in time (and thus the same initial prices for all
plain vanilla options such as European calls). We may wonder
whether our lognormal mixture diffusion dynamics~(\ref{Ssdemix})
is the local volatility version of some stochastic volatility
model. The answer is affirmative and we introduce the related
model below.

Consider the following uncertain volatility model:
\begin{eqnarray}\label{uncertmix}  d S_t &=& r(t) S_t dt + \xi(t) S_t dW_t,\ \  S_0 = s_0 ,  \\ \nonumber
(t \mapsto \xi(t)) &=& \left\{ \begin{array}{c}
(t \mapsto  \nu_1(t)) \  \mbox{with probability}  \  \lambda_1, \\
\ldots  \\
(t \mapsto  \nu_m(t))  \  \mbox{with probability} \ \ \lambda_m,
\end{array}\right. \\ \nonumber
 \xi  \mbox{ independent of} &W& \mbox{and drawn at random at (an almost zero) time}\ \epsilon>0,
\end{eqnarray}
where all $\nu$'s are assumed to be regular enough and have a
common value in $[0,\epsilon]$, $\nu_i(t) = \bar{\nu}$ for all $i$
and $t\le \epsilon$. We assume $\xi$ to be independent of $W$ and
that the original probability space is large enough to allow for
such a $\xi$ (otherwise we may define $\xi$ on a different space
and then take the product space). Conditional on $\xi$, the
process $S$ is a geometric Brownian motion as in the Black-Scholes
model. Thanks to independence, it is easy to show that for $t>
\epsilon
> u$,
\begin{eqnarray*} Q\{S_t\in A| S_u = y\}
= \sum_{i=1}^m  \lambda_i Q\{S_t\in A| S_u = y, \xi = \nu_i\}
\end{eqnarray*}
so that in particular
\[ p_{S_t|S_u}(x; y)  = \sum_{i=1}^m  \lambda_i   p_{S_t|S_u,\xi}(x; y,\nu_i)  , \]
i.e. the {\em transition density} of our uncertain volatility
model is a mixture of lognormal transition densities, each
corresponding to a volatility function $\nu_i$. If we condition on
an instant $u>\epsilon$, including the information on which value
$\nu_i$ of $\xi$ has realized itself at time $\epsilon<u$, an
information that is contained in the path of $S$ up to
$u>\epsilon$, the transition density between $u$ and $t$ reduces
merely to a lognormal density characterized by the relevant
$\nu_i$.

By considering the case $u=0$, it is immediate to see that this
model has the same marginals as the lognormal mixture diffusion
seen earlier, although it leads to an incomplete market. Then at
the initial time $0$ it implies the same prices as the local
volatility version seen earlier. Hedging is thus different and
more complicated, and has to be based on additional hedging
instruments. But what is interesting now is that transition
densities are also known, not only marginals. This is not true for
the local volatility version seen earlier.

There is, however, a close relationship between the lognormal mixture diffusion dynamics~(\ref{Ssdemix})
and the uncertain volatility mixture dynamics~(\ref{uncertmix}).

\begin{proposition}
The lognormal mixture diffusion dynamics~(\ref{Ssdemix}) is the local volatility version of the
uncertain volatility mixture dynamics~(\ref{uncertmix}). The two models are linked by the relationship
\[   \sigma^2_{\mbox{\tiny mix}}(t,x) = E\{\xi(t)^2 | S_t = x\}. \]
\end{proposition}
\begin{proof}
The proof is immediate by resorting to a variant of Bayes' formula:
\begin{eqnarray*}
 E\{\xi(t)^2 | S_t = x\} = E[\xi(t)^2 \sum_{k=1}^m 1\{\xi = \nu_k\} | S_t = x]
= \sum_{k=1}^m  E[\xi(t)^2 \ 1\{\xi = \nu_k\} | S_t = x] \\
= \sum_{k=1}^m E[\xi(t)^2  | S_t = x, \xi = \nu_k ] Q\{ \xi =
\nu_k| S_t = x\} = \sum_{k=1}^m \nu^2_k(t)  Q\{ \xi = \nu_k | S_t
= x \} =  \sigma^2_{\mbox{\tiny mix}}(t,x)
\end{eqnarray*}
since, by Bayes' formula,
\[ Q\{ \xi = \nu_k | S_t = x\} = Q\{ S_t \in dx |\xi = \nu_k \} Q\{ \xi = \nu_k\} / Q\{ S_t \in dx\} = \Lambda_k(t,x) . \]
\end{proof}

\begin{remark} (Casting some light on the ``zero terminal correlation" result of Theorem~\ref{thterminalcorr})
The terminal correlation computed at time $0$ between the asset $S$ and its average variance
$\int_0^T \xi^2(t) dt$ is easily seen to be zero, due to independence of $\xi$ and $W$. The same property is shared
by the local volatility version, as was pointed out in Theorem~\ref{thterminalcorr}. Now we see that the local volatility
version maintains the decorrelation pattern between volatility and underlying asset that is so natural for its uncertain
volatility originator. The result of Theorem~\ref{thterminalcorr} looks less surprising in the light of this result for the
uncertain volatility version.
\end{remark}

\subsection{Evolution of the volatility smile in the two models}
We now look at another important feature of the lognormal mixture
diffusion that has not been investigated in our earlier papers. We
are interested in looking at the smile evolution in time as
implied by the model.

We investigate this matter numerically as follows. We consider a
set of parameters coming from a possible calibration of
model~(\ref{Ssdemix}) with $m=2$ to the foreign exchange market.
The initial foreign exchange rate as from february 10, 2003 is
$S_0=1.07$ US Dollars for 1 Euro. Calibration of the model to
market data provides us with the parameters (we assume the $\nu$'s
to be constant in time, except for a negligible initial
time-interval $[0,2\epsilon]$) \[ \lambda_1 = 0.9747, \ \lambda_2
= 0.0253, \  \nu_1(t)= 0.7572, \  \nu_2(t)= 0.0899. \] The
risk-neutral drift $r(t)$ is taken consistently with the
differences of interest rates in the domestic and foreign curves:
indeed, we know that under the risk neutral measure, the drift of
an exchange rate is the difference of instantaneous interest rates
between the domestic and the foreign markets, $r(t) := r_d(t)-
r_f(t)$, where interest rates are assumed to be deterministic and
$r_d, r_f, R_d, R_f$ denote respectively the domestic and foreign
instantaneous interest rates and their integrals. The initial
smile for an option maturing in one year, $K\mapsto V(1y,K)$ is
given in the $0y$-row of Table~\ref{tablefuturesmile}.
We perform the following test. We consider options with one-year
time to maturity $T-t=1y$ set  at future times $t>0$, conditional
on the average underlying being realized, i.e. conditional on $S_t
= \bar{S}_t := E_0(S_t)$, the expectation taken under the risk
neutral measure. We therefore price the option with our
model~(\ref{Ssdemix}), by resorting to an Euler scheme with time
step of 1/1000y (Antignani~(2003)) and inverting then the
corresponding Black Scholes formula. Indeed, let us define
$V(t,T,K)$ as the solution of the equation
\begin{equation} \label{forwardsmiledef}
 e^{-R_d(t,T)} E_t[(S_T-K)^+|S_t =\bar{S}_t]  =
\mbox{BSCall}(\bar{S}_t,K,T-t,R_d(t,T),R_f(t,T),V(t,T,K)).
\end{equation}
By solving the above equation we infer what we call the
conditional future smile at $t$ for maturity $T$, i.e. $K \mapsto
V(t,T,K)/\sqrt{T-t}$. We will take $t=1y,2y,3y,6y,7y$ and
$T=t+1y$. We thus focus on options maturing in one year and on
their smile as implied by model~(\ref{Ssdemix}), which has been
employed to compute the expectation on the left hand side
of~(\ref{forwardsmiledef}). Interest rates are taken from the
market and assumed to be deterministic. Table~\ref{tablerates}
reports the market values we used. We recall that in a
deterministic interest-rates world $R(t,T) = R(0,T) - R(0,t)$.

\begin{table}[h]
\begin{center}
\begin{tabular}{|r|r|r|}
\hline
         $T$  &   $e^{-R_d(0,T)}$ & $e^{-R_f(0,T)}$   \\  \hline

         1y &   0.974454 &   0.985738 \\

         2y &   0.946724 &   0.960891 \\

         3y &   0.914757 &   0.925555 \\

         4y &   0.879548 &   0.885228 \\

         5y &   0.841922 &    0.84227 \\

         6y &   0.803363 &   0.799019 \\

         7y &   0.764915 &   0.756466 \\ \hline

\end{tabular}
\end{center}
\caption{Domestic and foreign discount factors (Euro and
USD)}\label{tablerates}
\end{table}

We obtain the results reported in Table~\ref{tablefuturesmile} as
annualized percentage implied volatilities. We notice that the
one-year smile flattens considerably in time (and similar
considerations apply to longer maturity smiles). The initial smile
(first row) shows an excursion from $16.17\%$ to $10.72\%$ and
then up to $13.61\%$, whereas the last  smile (7y/8y smile, last
row) moves from $10.26\%$ down to $9.55\%$ and up again to
$9.77\%$. As we see by looking at the rows of the table, the smile
flattens considerably in time.

\begin{table}[h]
\begin{center}
\begin{tabular}{|r|r|rrrrrrrrr|}
\hline & & & & & & $K/\bar{S}_t$ & & & & \\
         $t$ & $\bar{S}_t$ &  0.8 &       0.85 &        0.9 &       0.95 &          1 &       1.05 &        1.1 &       1.15 &        1.2 \\
\hline
        0y &  1.07 &    16.17 &      13.73 &      11.98 &      11.02 &      10.72 &      10.84 &      11.30 &      12.26 &      13.61 \\

        1y &  1.08257 &   11.23 &      10.44 &      10.10 &       9.98 &       9.96 &      10.00 &      10.10 &      10.30 &      10.62 \\

        2y &  1.08628 &    10.10 &       9.95 &       9.86 &       9.83 &       9.83 &       9.88 &       9.97 &      10.07 &      10.24 \\

        3y &  1.08294 &   10.60 &      10.09 &       9.92 &       9.83 &       9.81 &       9.82 &       9.85 &       9.91 &      10.02 \\

        6y &  1.06401 &   11.14 &      10.20 &       9.84 &       9.71 &       9.68 &       9.66 &       9.67 &       9.72 &       9.85 \\

        7y &  1.05773 &    10.26 &       9.83 &       9.67 &       9.60 &       9.57 &       9.55 &       9.57 &       9.64 &       9.77 \\
        \hline
\end{tabular}
\end{center}
\caption{Conditional future 1y smile $K \mapsto V(t,t+1y,K)$ for
$t=0,1,2,3,6,7$ years}\label{tablefuturesmile}
\end{table}

Again, the one-dimensional diffusion model~(\ref{Ssdemix}) mimics
the uncertain volatility model~(\ref{uncertmix}) of which it is
the local volatility version. Indeed, since after time $\epsilon$
we know the realization of $\xi$, we know which volatility $\nu_i$
realized itself and, conditional on this, our
process~(\ref{uncertmix}) is a geometric Brownian motion and
implies a flat smile. Therefore, after time $\epsilon$ the smile
flattens completely around the realized $\xi$ in the
model~(\ref{uncertmix}), and the one-dimensional diffusion version
mimics this behaviour by progressively flattening the implied
smile in time.

\section{Conclusions and further research}
In the present paper, we have found a candidate diffusion process
whose marginal law follows a given evolving mixture of probability
densities. We derived a stochastic differential equation (SDE)
admitting a unique strong solution whose density evolves as a
mixture of Gaussian densities. We introduced a seemingly
paradoxical result on the comparison between the instantaneous and
the terminal correlation between the obtained process and its
time-averaged squared diffusion coefficient. As an application to
option pricing, we considered diffusion processes whose marginal
densities are mixtures of lognormal densities, showing how such
processes can be used to model the market smile phenomenon.
Furthermore, we have pointed out how the lognormal mixture
dynamics is the one-dimensional diffusion version of a geometric
Brownian motion with uncertain volatility, leading to an uncertain
volatility Black-Scholes model, which allowed us to suitably
reinterpret the earlier correlation result. Finally, we checked
numerically the evolution of the smile in time and found that the
diffusion model mimics again the uncertain volatility model by a
substantial flattening of the smile implied curve in time.



\end{document}